# Diamond Processing by Focused Ion Beam– Surface Damage and Recovery

I. Bayn[1], A. Bolker[2], C. Cytermann[2], B. Meyler[1], V. Richter[2], J. Salzman[1] and R. Kalish[2]

1 Department of Electrical Engineering and Microelectronics Research Center, Technion Haifa, 32000, Israel

2 Department of Physics and Solid State Institute, Technion Haifa, 32000, Israel

The Nitrogen Vacancy color center ($NV^-$) in diamond is of great interest for novel photonic applications. Diamond nano-photonic structures are often implemented using Focused-Ion-Beam (FIB) processing, leaving a damaged surface which has a detrimental effect on the color center luminescence. The FIB processing effect on single crystal diamond surfaces and their photonic properties is studied by Time of Flight Secondary Ion Mass Spectrometry (TOF-SIMS) and photoluminescence (PL). Exposing the processed surface to hydrogen plasma, followed by chemical etching, drastically decreases implanted Ga concentration, resulting in a recovery of the $NV^-$ photo-emission and in a significant increase of the $NV^-/NV^0$ ratio.

The negatively charged nitrogen-vacancy color center (NV⁻) in diamond is an attractive candidate as a solid state quantum bit (qubit)[1]. Quantum information encoded into the NV⁻ ground state triplet allows implementation of qubit gates[2] and quantum registers[3], however for many qubit computation a photonic architecture for controlled distant spin-spin interaction is desirable. Single NV⁻ registered into high-$Q$ cavities coupled by waveguides enables realization of a photonic module[4] needed for three-dimensional cluster state based quantum computing[5]. The first diamond based photonic crystal (PC) nanocavity suitable for coupling of photons from the NV⁻ was realized in a nanodiamond (ND) film[6]. It would, however, be advantageous to realize these in single crystal diamond due to potentially lower losses[6]. The commonly used method of micro-machining and patterning diamond is by Focused-Ion-Beam (FIB) processing[7-9]. However this is accompanied by Ga implantation beneath the carved surface and by some "spill over" of the eroded material. As a result, the luminescence of FIB processed diamond is found to be strongly quenched, and a charge conversion of NV⁻ to NV⁰ is observed[7,9].

In this work, we report on the beneficial effect that hydrogen plasma followed by acid treatments has on the optical and luminescence properties of the FIB processed diamond surfaces. In contrast to the previous studies of plasma treatment[10,11] that leaves an H terminated surface, the acidic etching following the hydrogen plasma treatment renders the surface oxygen terminated, which has been shown to enhance the formation of NV⁻.[10]

The present study is conducted on FIB-milled type Ib diamond surfaces and on triangular cross section nanobridges realized in them. In order to allow accurate measurements of the thickness of the removed layer by the various employed processes, a deep Li layer (*40keV, $10^{15}ions/cm^2$*), was implanted into the blank diamond surface to serve as a depth marker[12]. Two square wells *200μ×200μ* and *95nm* deep were FIB milled using a *21nA, 30 keV* FIB beam. The first well served as reference for the Ga and Li as received depth profiles. The second well was exposed to hydrogen plasma (H$_2$ flow: *210sccm*, pressure: *60Torr,* MW power: *850W,* temperature: *650$^0$C*) for *80min*. Following each stage the sample was exposed to a boiling HClO$_4$: HNO$_3$:H2SO$_4$ (1:1:1) acid mixture. Similar treatments were done to the FIB milled triangular nanobridges discussed below.

In order to assess the effect of the various treatments on the properties of diamond, depth profiles were recorded at different stages of the treatment, using an ION-TOF TOF-SIMS 5 instrument. Fig. 1 displays (i) the as implanted Li depth profile prior to Ga milling (ii) the Li and Ga depth profiles following FIB milling and acid treatments (first well area) (iii) the Li and Ga depth profiles following a subsequent H plasma (80 minutes) and acid treatments (second well area). The data shown in Fig. 1 for the different probed sections were shifted in depth to take into account the surface milling/etching) so as to overlap the three identical Li profiles. These data show that the FIB treatment has removed *95 (±3) nm* from the diamond surface and that the subsequent H plasma and acid treatment has resulted in the removal of another *20 (±3) nm* in the Ga milled area. The measured Ga depth profile obtained after FIB milling (Fig. 1, red line, "Ga-init") fits very well the SRIM simulation[13] (Fig. 1, black dots, "Ga-SRIM"), taking into account the diamond material removal during the Ga ion bombardment. Following the 80 minute plasma treatment and etching, practically all Ga in the diamond surface has been removed (Fig. 1, blue line, "Ga-plsm"). To check the influence that the plasma and acid etching has on the surface properties two small wells *5μ×5μ* and *100nm* deep were milled by FIB using a very low beam current (*48pA*) to achieve good surface quality. These wells were characterized by atomic-force-microscopy (AFM) prior and after the treatment. The surface roughness (RMS) was found to decrease, following the treatment by *25%*.

A major question is whether the above described diamond treatments result in an improvement of photo emission of the NV centers and of light collection in diamond photonic devices. To investigate this we have studied the emission properties by micro-photoluminescence (μ-PL) in a series of nine triangular cross section nanobridges fabricated by FIB with different surface/volume ratios by varying the width (*W*) from *0.5μ* to *2μ* keeping length at *12μ* (see Fig. 2). The FIB conditions were: *30keV* Ga ions with beam currents gradually decreasing from *2.8nA* (for initial fast milling) to *98pA* (for final surface polishing)[9]. The bridges are separated by ~*3μ* from the substrate and by more than *5μ* from the side walls. μ-PL spectra were measured at room-temperature.[9]

The PL data for three diamond bridges with different widths (*0.5*, *1.3*, and *2.0 μ*) and that from a bulk (untreated) Ib diamond are shown in Fig.3, before (Fig. 3a (top)) and after H plasma and acid treatments (Fig 3a bottom). As can be seen, prior to the treatment the PL intensity in the bridge with *W=2μ* is ~*7* times lower than that of the unprocessed diamond. The spectrum contains several peaks (at *572, 578.4, 617.4, 638.5nm*). The *572nm* peak is the diamond Raman line. The peak at *578.4nm* (denoted NV$^*$ below) seems to be related to the NV$^0$ but is red-shifted by ~*3.4nm* with respect to the NV$^0$ zero-phonon-line (ZPL). The peak at *638.5nm* is the NV$^-$ ZPL. The origin of the peak at *617.4nm* requires further investigation, however, it is of no major relevance to the present discussion. The integrated peak areas of the NV$^*$ and NV$^-$ PL lines for the different bridges are shown in Fig. 3b (top). The NV$^*$ intensity exhibits an approximate linear dependence on the bridge width, suggesting that the center is formed in the vicinity of the treated surfaces. The maximum ratio of NV$^-$/NV$^*$≈*0.04* (for *W=2μ*) is ~$10^2$ *lower* than the NV$^-$/NV$^0$ in unprocessed bulk diamond. The NV$^-$ disappearance in the narrowest bridges (*W<1μ*) suggests that implanted Ga neutralizes NV$^-$ and promotes NV$^*$ formation. Increasing *W* beyond *1μ*, apparently, leaves the NV¯ in the bridge center nearly unaffected by the residual Ga on the milled walls.

The PL spectra from the same bridges following the H plasma and acid treatments are shown in the lower part of Fig 3a (bottom). The line shapes resemble those of bulk diamond, while the maximum intensity is *increased* by up to ~ *620* ($W=2\mu$) as compared to the pre-treatment values. Only $NV^0$ at *574.9nm ($\Delta\lambda=4.41nm$)* and the NV¯ at *638nm ($\Delta\lambda=3.16nm$)* peaks can be noticed. For the thick bridge ($W=2\mu$) the abundance of NV¯ is increased by $\sim 2.1 \times 10^3$ and NV¯/$NV^0$ ratio is ~*3.6*. In the PL spectrum of the thin bridge ($W=0.5\mu$) the NV⁻ signal can not be observed prior to the treatment, whereas following the H plasma and acid treatments the NV⁻ line is large (*$1.8 \times 10^4$* counts/s) and NV¯/$NV^0$=*2.36*. The NV¯ and $NV^0$ abundance increases linearly with bridge cross-section (Fig 3b (bottom)), indicating that the signal originates from the entire bridge volume and not from its carved surface only. The total increase in PL is by the factor of ~*$10^2$* as compared to the bulk. This increase cannot be explained by Ga removal or by $NV^*$ conversion to $NV^0$ and to NV¯ only; however since each Ga ion produces approximately *170* vacancies/ion (SRIM), we suggest that after chemical treatment these vacancies are converted to NVs, preferentially to NV¯.

In summary, it was shown that whereas FIB processing of diamond degrades the optical characteristics, exposure of such treated surfaces to a hydrogen plasma and to acid cleaning greatly improves the PL of color centers in diamond photonic structures. TOF-SIMS depth profiling shows that the detrimental effect of FIB is mainly related to the presence of implanted Ga at or near the surface region. Once the Ga rich layer is removed by H plasma, and the thus formed H termination is replaced with O termination by exposure to acids, a drastical improvement in the optical properties is noticed. Furthermore, this treatment enhances the desirable NV¯ emission by a factor of *~50* with respect to unprocessed diamond. These results open new ways for realization of FIB-based diamond photonic device with improved photonic properties.

This work was supported by the Russell Berrie Nanotechnology Institute at the Technion and by the German Israeli Foundation (GIF) contract # I-1026-9.14/2009.

**Figures Captions:**

1. FIG. 1. TOF-SIMS Ga and Li depth profile insde and outside *95nm* deep wells FIB milled in diamond, before and after hydrogen plasma and chemical etching. The diamond was implanted by 40keV $Li^+$ as a depth marker prior to the milling. i) as implanted Li depth profile before FIB milling (green curve) ii) Li and Ga depth profiles obtained in the FIB milled well post chemical etching (cyan and red curves, respectively) iii) Li and Ga depth profiles obtained in the FIB milled well, after H plasma and chemical etching (pink and blue curves, respectively) iiii) SRIM simulation of the FIB implanted Ga, fitting exactly the measured profile (Black dotted curve). The figure shows ~*20nm* damaged diamond surface etching after treatments. Note, the Li profiles coincide accurately but are slightly scaled down for an easier picture reading.

2. FIG. 2. Scanning electron micrograph (SEM) of *2μ* wide bridge carved with FIB.

3. FIG 3. (a) The PL of there bridges (*W=0.5, 1.3, 2μ*) carved by FIB (left *y*-axis) and bulk unprocessed diamond as a reference (--, right *y*-axis). (b) $NV^*$ and $NV^-$ integral PL (background subtracted and Gaussian fit) that is proportional to their abundance. The *W<1μ* is marked in cyan. (c) The PL of bridges measured in (a) post hydrogenation and chemical milling. (d) Integral intensity of $NV^0$ and $NV^-$ post hydrogenation and chemical milling.

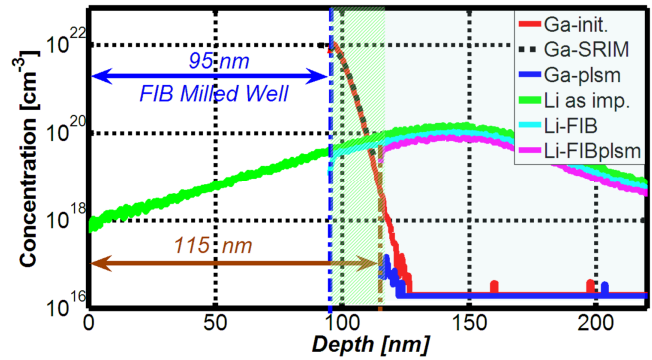

FIG. 1. TOF-SIMS Ga and Li depth profile insde and outside *95nm* deep wells FIB milled in diamond, before and after hydrogen plasma and chemical etching. The diamond was implanted by 40keV Li$^+$ as a depth marker prior to the milling. i) as implanted Li depth profile before FIB milling (green curve) ii) Li and Ga depth profiles obtained in the FIB milled well post chemical etching (cyan and red curves, respectively)  iii) Li and Ga depth profiles obtained in the FIB milled well, after H plasma and chemical etching (pink and blue curves, respectively) iiii) SRIM simulation of the FIB implanted Ga, fitting exactly the measured profile (Black dotted curve). The figure shows ~*20nm* damaged diamond surface etching after treatments. Note, the Li profiles coincide accurately but are slightly scaled down for an easier picture reading.

Figure 1 of 3

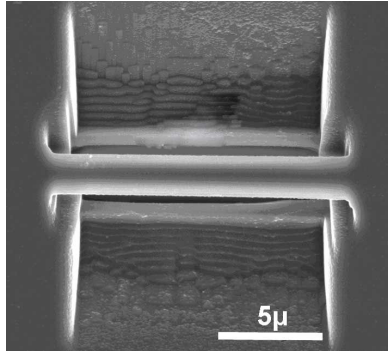

FIG. 2. Scanning electron micrograph (SEM) of *2μ* wide bridge carved with FIB.

Figure 2 of 3

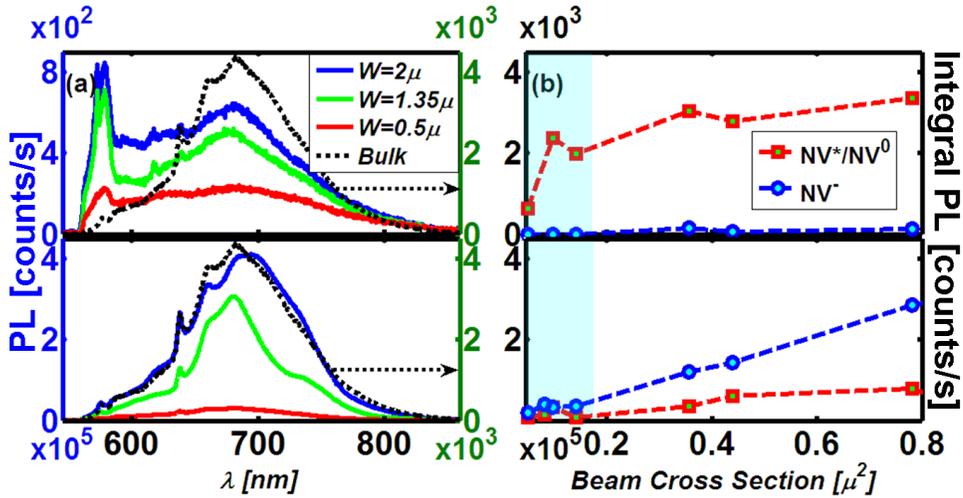

FIG 3. (a) The PL of there bridges ($W=0.5, 1.3, 2\mu$) carved by FIB (left $y$-axis) and bulk unprocessed diamond as a reference (--, right $y$-axis). (b) $NV^*$ and $NV^-$ integral PL (background subtracted and Gaussian fit) that is proportional to their abundance. The $W<1\mu$ is marked in cyan. (c) The PL of bridges measured in (a) post hydrogenation and chemical milling. (d) Integral intensity of $NV^0$ and $NV^-$ post hydrogenation and chemical milling.

Figure 3 of 3